\documentclass[12pt]{article}
\usepackage{graphicx}
\title{Stabilities of one-dimensional stationary states of Bose-Einstein condensates}
\author{Wenhua Hai$^{1}$\footnote{Author to whom
all correspondence should be addressed, Email address:
whhai2005@yahoo.com.cn}, \ Guishu Chong$^{1,2}$,\ Jianwen Song$^{1}$
\\ $^{1}$Department of Physics, Hunan Normal University,  and \\
Key Laboratory of Quantum Structure and Manipulation of
\\Ministry of
Education, Changsha 410081, China \\ $^{2}$Department of physics,
Hunan University, Changsha 410082, China}
\date{}
\begin{document}

\maketitle
\begin{abstract}

We explore the dynamical stabilities of a quasi-one dimensional (1D)
Bose-Einstein condensate (BEC) consisting of fixed $N$ atoms with
time-independent external potential. For the stationary states with
zero flow density the general solution of the perturbed time
evolution equation is constructed, and the stability criterions
concerning the initial conditions and system parameters are
established. Taking the lattice potential case as an example, the
stability and instability regions on the parameter space are found.
The results suggest a method for selecting experimental parameters
and adjusting initial conditions to suppress the instabilities.



\end{abstract}

\section{Introduction}

Experimental observation of atomic gas Bose-Einstein condensates
(BECs) has caused significant stimulation to the study of
macroscopic quantum phenomena with nonlinearity. In the mean field
regime where the BECs are governed by the Gross-Pitaevskii equations
(GPE), the BEC of a stationary state can be observed carefully in
experiments only for the stable solutions of GPE. For the purpose of
applications, the studies on the stability and instability of the
solutions of GPE are necessary and important
\cite{Chin}-\cite{Saito}. Recently, the instabilities of BECs have
attracted much interest and the corresponding experimental
\cite{Chin, Fallani, Burger} and theoretical
\cite{Zheng}-\cite{FAbdullaev} works were reported for various BEC
systems. Several different definitions such as the Landau
instability \cite{Wu, Machholm}, dynamical instability
\cite{Bronski}, quantum instability \cite{Shchesnovich}, parametric
instability \cite{Genkin} and modulational instability
\cite{Konotop} were employed. The used research methods concerned
the characteristic exponent technique \cite{Zheng}, Gaussian
variational approach \cite{Abdullaev}, and the numerical simulations
to the partial differential equations \cite{Wu, Bronski, Deconinck}.
The reported results showed that the instabilities are associated
with the BEC collapse \cite{Konotop, Kagan}, implosion and chaos
\cite{Saito2} - \cite{Xia}, dynamical superfluid-insulator
transition \cite{Smerzi}, and the formation and evolution of the
matter-wave bright solitons \cite{Strecker, Carr, Salasnich}. In
order to stabilize the BECs \cite{Saito}, some stability criteria
\cite{Berge} and parameter regions \cite{Zheng, Wu, Luo, Montina}
were demonstrated. Most of the works focus in the stabilities under
random perturbations. Experimentally \cite{Burger} and theoretically
\cite{Wu} investigating the stabilities under the controllable
perturbations has also become a challenging problem.

In the sense of Lyapunov, the instability entails that the initially
small deviations from the unperturbed state grow without upper
limit. We shall restrict the dynamical instability to the particular
case of nonzero characteristic exponents such that the minor
deviations from the unperturbed state grow exponentially fast
\cite{Wu, Bronski}. All of the above-mentioned investigations on the
dynamical stabilities and instabilities are based on such a type of
instability. By the control of instability we mean to induce the
transitions from unstable states to stable ones. Realization of the
control needs selecting the system parameters to enter the stability
regions, or initially using a controllable perturbation as a control
signal to suppress the growth of perturbed solutions. Any experiment
always contains a degree of noise, that leads to the random
perturbations to the system. Therefore, in order to suppress the
known unstable motions, we have to initially adjust the system by
using the control signal being stronger than the noise.

In the previous work, we have investigated the stabilities of BECs
for the time-dependent chaotic states \cite{whai, Chong} and
dissipative cases \cite{Luo}. In this paper, we shall consider the
dynamical stability of the stationary states for a quasi-1D BEC
consisting of fixed $N$ atoms with time-independent external
potential and atomic scattering length. It will be demonstrated that
for the case of zero flow density the bounded perturbed solutions
depend on the external potential, condensed atom number, and the
initial disturbances. The dependence implies that the stationary
state of BEC is certainly stable only for the given parameter region
and the possible instability can be suppressed by some initial
adjustments. We take the BECs held in an optical lattice as an
exemplification to illustrate the results on the stability,
instability and undetermined stability. The results contain the
known analytical assertions for the optical potential case \cite{Wu,
Bronski} and supply a method for selecting experimental parameters
and adjusting initial conditions to establish the stable motions of
BEC.

\section{Linearized equations and their solutions in the case of zero flow density}
We start with the dimensionless quasi-1D GPE \cite{Bronski,Dalfovo,
Leggett}
\begin{eqnarray}
i \psi_t=- \frac 1 2 \psi_{xx} + [V(x) +g_{1}|\psi|^2]\psi,
\end{eqnarray}
where the suitable units with $\hbar=m=1$ have been considered,
$V(x)$ denotes the external potential, the quasi-1D interaction
intensity $g_{1}$ is related to the $s$-wave scattering length
$a_s$, atomic mass $m$ and the transverse trap frequency $\omega_r$
\cite{Gardiner, Hai2} for the normalized wave-function $\psi$ with
norm $|\psi|^2$ being the linear density of atomic number
\cite{Bronski, Leggett}. It is well known that different solutions
of a nonlinear equation may possess different stabilities. Here we
study stability only for the stationary state solution of the form
\begin{eqnarray}
\psi_0=R(x)\exp [i\theta(x)-i\mu t],
\end{eqnarray}
where $\mu$ is the chemical potential, $R(x)$ and $\theta(x)$
represent the module and phase, which are both the real functions.
In the considered units, the phase gradient $\theta_x$ is equal to
the flow velocity field. Given the module, we define the useful
Hermitian operators \cite{Bronski}
\begin{eqnarray}
L_n=-\frac 1 2 \frac{\partial ^2}{\partial x ^2}+n g_1 R^2+V(x)-\mu,
\ \ \ for \ \ n=1,3.
\end{eqnarray}
Then inserting Eq. (2) into Eq. (1) gives the equations
\begin{eqnarray}
L_1 R(x)=0.
\end{eqnarray}
In the equation we have assumed the flow velocity field and current
density being zero.

We now investigate the stability of stationary state Eq. (2) by
using the linear stability analysis, which is associated with
boundedness of the perturbed solution \cite{Bronski, Berge}
\begin{eqnarray}
\psi=[R(x)+\varepsilon \phi_1(x,t)+i \varepsilon \phi_2(x,t)]\exp
[i\theta(x)-i\mu t],
\end{eqnarray}
where the perturbed correction $\varepsilon\phi_i(x,t)$ is real
function with constant $|\varepsilon|\ll 1$. Substituting Eqs. (5)
and (4) into Eq. (1) yields the linearized equations
\cite{Bronski}
\begin{eqnarray}
\phi_{1t}=L_1 \phi_2, \ \ \ \ \ \phi_{2t}=-L_3\phi_1.
\end{eqnarray}
For most of external potentials $V(x)$ we cannot derive the exact
solutions from Eq. (1) or Eq. (4) such that the operators $L_n$
cannot be determined exactly. In the case of optic lattice
potential, some specially exact solutions have been found
\cite{Bronski, Deconinck, Hai2}, however, solving Eq. (6) for the
general solution is still difficult. Therefore, we have to focus our
attentions to the dynamical stability which is associated with the
perturbed solutions of space-time separation,
\begin{eqnarray}
\phi_i(x,t)=T_i(t)\varphi_i(x), \ \ \ for \ \ \ i=1,2.
\end{eqnarray}
Note that the real function $\phi_i$ limits $T_i$ and $\varphi_i$ to
real or imaginary simultaneously, the difference between both is
only a sign $``-"$ of $\phi_i$. We take real $T_1, \varphi_1, T_2$,
and $\varphi_2$ without loss of generality, since the changes of the
signs of $\phi_i$ do not affect the stability analysis. We shall
discuss how to establish the sufficient conditions of stability as
follows.

Combining Eq. (6) with Eq. (7), we get the coupled ordinary
differential equations
\begin{eqnarray}
\dot T_1(t)&=&\lambda_1 T_2(t), \ \ \dot T_2(t)=-\lambda_2 T_1(t);
\\  L_3 \varphi_1(x)&=&\lambda_2 \varphi_2(x), \ \  L_1
\varphi_2(x)=\lambda_1 \varphi_1(x).
\end{eqnarray}
Here $\lambda_i$ is the real eigenvalue determined by the initial
perturbations $\dot T_i(0),T_i(0)$. The corresponding decoupled
equations are derived easily from the coupled ones as
\begin{eqnarray}
\ddot T_i(t)=&-& \lambda_1 \lambda_2 T_i(t), \lambda_1=\frac{\dot
T_1(0)}{T_2(0)}, \lambda_2=-\frac{\dot
T_2(0)}{T_1(0)}; \\
L_1 L_3 \varphi_1&=&\lambda_1 \lambda_2 \varphi_1, \ \ \ \ L_3 L_1
\varphi_2=\lambda_1 \lambda_2 \varphi_2.
\end{eqnarray}
Obviously, the general solutions of Eq. (10) can be written as the
exponential functions
\begin{eqnarray}
T_i=A_ie^{\lambda t}+B_i e^{-\lambda t}, \ \ \
\lambda=\sqrt{-\lambda_1 \lambda_2}, \ \ \ \ \ \ \ \ \ \ \ \nonumber \\
A_i=\frac 1 2 \Big[T_i(0)+\frac{1}{\lambda}\dot T_i(0)\Big], \
B_i=\frac 1 2 \Big[T_i(0)-\frac{1}{\lambda}\dot T_i(0)\Big],
\end{eqnarray}
where $A_i, B_i$ are real or complex constants, which make $T_i(t)$
the real functions. Based on the existence of bounded eigenstates
$\varphi_i(x)$, the results are classified as the three cases:

(!`) {\bf Stability criterion}: The eigenstates of Eq. (11) are
bounded if and only if their eigenvalues are positive, $\lambda_1
\lambda_2=-\lambda^2>0$, that makes $\lambda$ the imaginary constant
and $T_i$ the periodic functions.

(!`!`) {\bf Instability criterion}: One can find a negative
eigenvalue $\lambda_1 \lambda_2=-\lambda^2<0$ associated with a set
of bounded eigenstates of Eq. (11) that makes $T_i$ the real
exponential function.

(!`!`!`) {\bf Undetermined stability}: One cannot determine whether
all eigenvalues of the bounded eigenstates of Eq. (11) are positive.
In this case, we can use criterion  (!`) to control the possible
instability of case (!`!`).

\section{Stability regions on the parameter space and control of instability}

It is interesting noting that if the initial perturbations can be
determined, the dynamical instability of real $\lambda$ case can be
controlled by adjusting the initial disturbances to obey $A_i=0$
that will suppress the exponentially rapid growth of $T_i$ in Eq.
(12). From Eqs. (12) and (10) we establish the controlling criteria
for the instability as $ \dot T_i(0)=- \lambda T_i(0).$ However, for
the random initial perturbations such a control is difficult to do,
since we cannot determine the initial values $ \dot T_i(0)$ and $
T_i(0).$ Therefore, in the case of random perturbation we are
interested in determining the same eigenvalue $\lambda_1 \lambda_2$
of operators $L_1 L_3$ and $L_3 L_1$, since the stability can be
established if and only if the eigenvalue is positive such that
$\lambda^2=-\lambda_1 \lambda_2<0$. Let $\alpha\ge \alpha_g$ and
$\beta\ge \beta_g$ be the eigenvalues of operators $L_1$ and $L_3$,
which are determined by the eigenequations $L_1 u(x)=\alpha u(x), \
\ L_3v(x)=\beta v(x)$ with $u$ and $v$ being their eigenfunctions,
where $\alpha_g$ and $\beta_g$ express the corresponding ground
state eigenvalues respectively. From Eq. (3) we know the relation
$L_3= L_1+2g_1 R^2$ that means $\alpha_g <\beta_g$ for $g_1>0$ and
$\alpha_g>\beta_g$ for $g_1<0$. It is clear that Eq. (4) is one of
the eigenequations of $L_1$ for the eigenvalue $\alpha=0$ so that
the ground state eigenvalue obeys $\alpha_g\le 0$ for any $g_1$.
Then $\beta_g$ can be positive or negative for $g_1>0$ and
$\beta_g<0$ for $g_1<0$.

From the above-mentioned results we establish the stability and
instability conditions:

Case $g_1>0$: The sufficient condition of stability is $\alpha_g =
0$, since such a ground state eigenvalue implies $\alpha \ge 0$ and
$\beta> 0$ for all of the eigenstates such that the well known
spectral theorem gives \cite{Bronski, Deconinck, Courant} $\lambda_1
\lambda_2\ge 0$. The corresponding sufficient conditions of
instability reads $\alpha_g < 0$ and $\beta_g> 0$.

Case $g_1<0$: The ground state eigenvalues satisfy the inequality
$\beta_g<\alpha_g\le 0$. So the sufficient condition of instability
is $\alpha_g= 0$.

In all the other cases, we don't know whether $\lambda_1 \lambda_2$
is certainly positive or negative, so the linear stabilities are
analytically undetermined. It is worth noting that Eq. (4) infers
$R(x)$ to be one of the eigenstates of $L_1$ with eigenvalue
$\alpha_R=0$. Therefore, if $R(x)$ is a ground state, the above
stability and instability conditions indicate that this state is
stable for $g_1>0$, and unstable (or metastable) for $g_1<0$.

\bf Note that all the above-mentioned results are valid for
arbitrary time-independent potential. \rm We will take the BEC held
in an optical lattice as a concrete physical example to evidence
these results. \bf In the lattice potential case, \rm the
above-mentioned sufficient conditions agree with the stability and
instability criterions established by the authors of Ref.
\cite{Bronski}. We shall apply the sufficient stability and
instability conditions to find the corresponding stability and
instability regions on the parameter space, and apply these results
to study the stabilization of the considered BEC system.

For an arbitrary time-independent potential, the eigenequation $L_1
u=\alpha u$ can be rewritten as the integral form \cite{whai2}
\begin{eqnarray}
u&=&u_1+u_2, \nonumber \\ u_1&=&q^{-1}e^{-qx}\int e^{qx}f u dx, \
u_2=-q^{-1}e^{qx}\int e^{-qx}f u dx, \nonumber \\
f&=&q^2/2+\alpha+\mu-V(x)-g_1R^2(x).
\end{eqnarray}
where $q>0$ is a real constant. This integral equation can be
directly proved by taking the second derivative from its both sides.
The integrals in Eq. (13) are indefinite, what means that the
solutions are defined with accuracy of two additive constants. While
the eigenequation $L_1 u=\alpha u$ just is a second order equation
which also implies two arbitrary constants determined by the
boundary conditions. It is the two additive constants to make the
integral equation (13) completely equivalent to the eigenequation.
The stability requires the eigenstate to be bounded and the possible
bounded solution $u$ must satisfy the boundedness condition
$\lim_{x\rightarrow\pm\infty}\int e^{\mp qx}f udx=0$. Under this
condition and for the \bf lattice potential case,\rm we can apply
the l'H$\ddot{o}$pital rule to get the superior limit \cite{Chong}
\begin{eqnarray}
\overline {\lim_{x\rightarrow\pm\infty}}u\le
\overline{\lim_{x\rightarrow\pm\infty}}u_1+\overline{\lim_{x\rightarrow\pm\infty}}u_2
=2q^{-2}\overline{\lim_{x\rightarrow\pm\infty}}(f u).
\end{eqnarray}
Note that there is not the usual limit, because of the periodicity
of lattice potential. It is clear that the solution of linear
equation $L_1 u=\alpha u$ can be taken as $u(x)=Au'(x)$ with
arbitrary constant $A$ and any solution $u'(x)$ such that one can
always select $u$ to obey
$\overline{\lim}_{x\rightarrow\pm\infty}u>0$. Thus Eq. (14) implies
$2q^{-2}\overline{\lim}_{x\rightarrow\pm\infty}f\ge 1$, namely
\begin{eqnarray}
\alpha\ge
-\{\mu+\overline{\lim_{x\rightarrow\pm\infty}}[-V(x)-g_1R^2(x)]\}=\alpha_g.
\end{eqnarray}
For the eigenequation $L_3 v=\beta v$ after using $3g_1$ instead of
$g_1$, the same calculations give
\begin{eqnarray}
\beta\ge
-\{\mu+\overline{\lim_{x\rightarrow\pm\infty}}[-V(x)-3g_1R^2(x)]\}=\beta_g.
\end{eqnarray}
Combining Eq. (15) with the stability sufficient condition
\cite{Bronski} $\alpha_g= 0$ for $g_1>0$, we get the parameter
region of stability
\begin{eqnarray}
\mu=\mu_s = -\overline{\lim_{x\rightarrow\pm\infty}}\
[-V(x)-g_1R^2(x)] \ \ \ for \ \ g_1>0,
\end{eqnarray}
which contains the relation among $\mu, g_1$ and the potential
parameters. Applying Eqs. (15) and (16) to the instability
sufficient conditions \cite{Bronski} $\alpha_g< 0,\ \beta_g> 0$ for
$g_1>0$ and $\alpha_g= 0$ for $g_1<0$, we get the parameter regions
of instability
\begin{eqnarray}
&-&\overline{\lim_{x\rightarrow\pm\infty}}\ [-V(x)-g_1R^2(x)]<
\mu=\mu_{in} < -\overline{\lim_{x\rightarrow\pm\infty}}\
[-V(x)-3g_1R^2(x)] \ \ for \  g_1>0;\nonumber \\
& & \mu_{in} = -\overline{\lim_{x\rightarrow\pm\infty}}\
[-V(x)-g_1R^2(x)] \ \ \ for \ \ g_1<0.
\end{eqnarray}
By the sufficient conditions we mean that the stationary state
$R(x)e^{-i\mu t}$ of Eq. (1) is certainly stable for the $\mu$
values in the parameter region fixed by Eq. (17), and the stationary
states are certainly unstable  for the $\mu$ values in any region of
Eq. (18). The dynamical stabilities are undetermined outside.

We now see the physical meaning of the stability relation in Eq.
(17) for the stationary states of BEC with zero current density.
Setting the sum of external potential and internal interaction as
$U(x)=V(x)+g_1R^2(x)$ with periodic $V(x)$ and bounded $R(x)$, when
$U(x)\ge B$ is satisfied for all $x$ values and a fixed constant
$B$, Eq. (17) implies $\mu_s= B\le U(x)$. Namely the sufficient
stability condition means that if the chemical potential is equal to
the minimum of $U(x)$, the considered states are certainly stable.
For a known state the stability can be easily examined by using Eq.
(17). We have tested the exact solutions given in Ref.
\cite{Bronski} for the potential $V(x)=-V_0 sn^2(x,k)$ and found
that some of them have the instabilities and undetermined
stabilities, where $|V_0|$ is the potential depth and $sn(x,k)$ the
Jacobian elliptic sine function with $k$ being the modulus.
Substituting one of the exact solutions,
$g_1R^2(x)=-(1+V_0/k^2)[1-k^2 sn^2(x,k)]$ with the potential depth
$-V_0\ge k^2$ and chemical potential $\mu=-1-V_0/k^2+k^2/2$ [see Eq.
(12) of Phys. Rev. E63, 036612(2001)], into Eq. (17) yields the
stability parameter relation $\mu_s=-1-V_0/k^2$. A difference of
$k^2/2$ exists between the $\mu_s$ value required by the stability
condition and the chemical potential $\mu$ in the exact solution,
namely the stability criterion (17) is not met here. This assertion
differs from the result of Ref. \cite{Bronski}, where this solution
fits their stability criterion and the stability is independent of
the parameters $k$ and $V_0$. However, when the potential depth
$|V_0|$ is much greater than the modulus $k$ (e.g. $V_0=-1$ and
$k=0.2$), we have the chemical potential near the stability relation
(17) ($ \mu=24.02=\mu_s+0.02\approx \mu_s$). This infers the higher
stability being associated with a smaller value of the modulus $k$
and a relatively greater $|V_0|$ value. Thus our stability parameter
criterion suggests that for a known solution with instability or
undetermined stability one can raise the practical stability by
adjusting the system parameter (e.g. the above $k$ and $|V_0|$) to
approach the values of the stability region in Eq. (17).

Generally, constructing a stable exact solution of GPE is not easy,
because of the non-integrability of Eq. (4) with periodic potential.
However, in the large-$N$ limit, we can fulfil the criterion (17)
for the case of a repulsive nonlinearity, since the Thomas-Fermi
(TF) approximation \cite{Dalfovo} $U(x)=\mu_{TF}$ just fits the
stability relation. Therefore, it is practical relevant to prepare
such a stable TF state $R(x,\mu_{TF})$ by increasing the condensed
atom number $N$. Given the number $N$ and the periodic boundary
condition experimentally, from the normalization condition
$N=n\int_0^{\pi} R^2(x,\mu_{TF})dx=n\int_0^{\pi}
[\mu_{TF}-V(x)]dx/g_1$ we derive the chemical potential of the
stable TF state
\begin{eqnarray}
\mu_{TF}=\mu_s=\frac{Ng_1}{n\pi}+\frac {1}{\pi}\int_0^{\pi} V(x)dx
\end{eqnarray}
which is related to the atom number $N$ and the potential strength
$V_0$ and period $K (k)$, where $n\sim 100$ is the lattice number.
In fact, noticing the dependence of $R=R(x,\mu)$ on $\mu$ in Eq.
(4), the normalization condition of any known state can also lead to
$\mu=\mu (N)$ and $R=R(x, N)$. Applying them to eliminate $\mu$ in
Eqs. (17) and (18) will give the corresponding relationships among
the experimental parameters $N, g_1,V_0$ and $K(k)$. So we can
control the instability of the known state by selecting the
experimental parameters to fit or to approach the stability region
of Eq. (17). In many practical cases, we cannot obtain the exact
solution of Eq. (4) for some periodic potentials, that necessitates
the numerical investigation. In order to fit (or near) the stability
region in Eq. (17) and to avoid the instability regions of Eq. (18),
we could use Eq. (19) to estimate and adjusted the chemical
potential in region $\mu\approx \mu_{TF}$ such that the stability of
the numerical solutions of Eq. (4) can also be established or
improved.

On the other hand, in the case of arbitrary time-independent
potential, for some known unstable solutions $R=R(x,\mu_{in})$ from
Eqs. (10) and (12) we can experimentally set and adjust the
initially controllable perturbation as a control signal
\cite{Burger} to suppress the exponentially fast growth of $T_i(t)$.
Although the phase $\theta$ and amplitude $R$ are time-independent
in the considered case, the initial perturbations can result in the
nontrivial and time-dependent corrections to the phase and
atomic-number density. From Eq. (5) we find their first corrections
as
\begin{eqnarray}
&&\triangle\theta(x,t) \approx \arctan [\varepsilon
T_2(t)\varphi_2(x)/R(x)]\approx \varepsilon T_2(t)\varphi_2(x)/R(x),
\nonumber \\ && \triangle |\psi|^2(x,t)\approx 2\varepsilon
T_1(t)\varphi_1(x)R(x),
\end{eqnarray}
which are initially proportional to $T_1(0)$ and $T_2(0)$
respectively. Making use of Eq. (20), the adjustments to the
initially controllable perturbations can be performed by trimming
the number density $|\psi|^2$, velocity field $(\triangle\theta)_x$
and their time derivatives which are proportional to the
corresponding trimming velocities. Given Eqs. (10) and (12), we know
the stability initial criterion
\begin{eqnarray}
\lambda^2=-\lambda_1 \lambda_2=\dot T_1(0)\dot
T_2(0)/[T_1(0)T_2(0)]<0.
\end{eqnarray}
Once Eq. (21) is satisfied in the adjustments to the initial
perturbations, Eq. (12) becomes the periodic solution which implies
the stability. Although we cannot determine the initial values $\dot
T_i(0)$ and $ T_i(0)$, experimentally, the number density can be
adjusted by varying the condensed atom number, and the adjustments
to superfluid velocity may be related to a displacement $\triangle
x$ of a magnetic potential \cite{Burger}. According to Eqs. (20) and
(21), if we initially increases (or decreases) both the relative
derivative $\dot T_2(0)/ T_2(0)=\frac{\partial
\triangle\theta_x(x,t)}{\partial t}|_{t=0}/\triangle\theta_x(x,0)$
of flow velocity and the relative derivative $\dot T_1(0)/
T_1(0)=\frac{\partial \triangle |\psi|^2(x,t)}{\partial
t}|_{t=0}/\triangle |\psi|^2(x,0)$ of atomic number density, the
stability initial criterion (21) is destroyed and the system will
become unstable. But when one of them is increased and another is
decreased simultaneously, the stability criterion (21) is satisfied
and the possible instability is suppressed. These assertions may be
tasted experimentally.

\section{Conclusions and discussions}

In conclusion, we have investigated the dynamical stability,
instability and undetermined stability of a quasi-1D BEC in the
stationary states for time-independent external potential and atomic
scattering length, and fixed atomic number. After space-time
separation of variables, we derive the general solutions of the
linearized time-evolution equations for the trivial phase case and
give a stability criterion related to the initial conditions. As an
important example, we evidence the stability criterion analytically
for the BEC held in an optical lattice potential. By using the known
sufficient conditions of stability and instability \cite{Bronski},
several parameter regions of stability and instability are shown.
Our results contain some new stability predictions which can be
tested with current experimental setups. Finally, we stress that
applying our stability initial criterion and parameter region one
can stabilize the considered BEC system by adjusting the system
parameters experimentally to enter or near the stability region of
Eq. (17) on the parameter space. For the parameters out of the
stability region we can also establish or improve the stability by
adjusting the initial flow velocity and atomic number density to fit
or approach the stability initial criterion.

\bf Acknowledgment \rm This work was supported by the National
Natural Science Foundation of China under Grant No. 10575034 and by
the Key Laboratory of Magnetic Resonance and Atomic and Molecular
Physics of China under Grant No. T152504.

{}

\end{document}